\documentclass[prc,preprint,showpacs,showkeys,preprintnumbers,amsmath,amssymb]{revtex4}

\usepackage{graphicx} 
\usepackage{dcolumn}  
\usepackage{bm}       

\begin{document}

\preprint{IKP DA}

\title{$^{187}$Re($\gamma$,n) cross section close to and above the neutron threshold}

\author{S. M\"{u}ller}
\email{mueller@ikp.tu-darmstadt.de}
\author{A. Kretschmer}
\author{K. Sonnabend}
\author{A. Zilges}
\author{D. Galaviz}
\altaffiliation[current address: ]{NSCL, Michigan State University, 
1 Cyclotron Lab, East Lansing, MI 48824-1321, USA}
\affiliation{Institut f\"{u}r Kernphysik, Technische Universit\"{a}t Darmstadt, 
Schlossgartenstrasse 9, D-64289 Darmstadt, Germany}

\date{\today} 

\begin{abstract}
The neutron capture cross section of the unstable nucleus $^{186}$Re 
is studied by investigating the inverse photodisintegration 
reaction $^{187}$Re($\gamma$,n). The special interest of the {\it s}-process 
branching point $^{186}$Re is related to the question of possible {\it s}-process 
contributions to the abundance of the {\it r}-process chronometer nucleus $^{187}$Re. 
We use the photoactivation technique to measure photodisintegration rates.  
Our experimental results are in good agreement with two different statistical model 
calculations. Although the cross sections predicted by both models for the inverse reaction
$^{186}$Re(n,$\gamma$) is too low to remove the overproduction of $^{186}$Os; the two
predicted neutron-capture cross sections differ by a factor of $2.4$; this calls for future 
theoretical study.
\end{abstract}


\maketitle

\section{\label{para:par1} Introduction}

Almost all elements above mass $A \approx 60$ can be produced in neutron capture 
reactions\,\cite{burb57}. Two different neutron induced processes are necessary 
to explain the abundance distribution of heavy elements. The first one is the 
slow neutron capture process ({\it s}-process). The neutron densities are of the 
order of $n_{\rm n} \approx 10^8\,{\rm cm}^{-3}$ and the time scale $\tau_{\rm n}$ 
between two subsequent neutron capture reactions is typically of the order of years. 
The {\it s}-process path propagates along the valley of stability. Whenever 
an unstable nucleus with a mean lifetime $\tau \ll \tau_{\rm n}$ is reached, this
nucleus $\beta$-decays. If $\tau \approx \tau_{\rm n}$,
a branching occurs and the {\it s}-process path splits. Thus, nuclei with 
$\tau \approx \tau_{\rm n}$ are called branching points of the {\it s}-process.
The second process is the rapid neutron capture process ({\it r}-process). High neutron 
densities ($n_{\rm n} \gg 10^{20} \,{\rm cm}^{-3}$) 
lead to the production of very neutron 
rich nuclei up to 20 mass units away from stable nuclei.
During freeze out, these  nuclei $\beta$-decay back to the valley of 
stability. 

There are at least two scenarios known where the {\it s}-process takes place. 
It occurs during helium burning in red giant stars and 
during helium shell flashes in low mass asymptotic giant branch (AGB) 
stars\,\cite{Kaep90, Arla99}. The former scenario is mainly responsible for 
the production of elements between iron and yttrium. The 
latter, for the production of elements between zirconium and 
bismuth. For a detailed discussion see e.g.\,\cite{Wall97}. 
In the following we will focus on the mass
region $A \approx 185$ and, hence, restrict our discussion to 
the so-called main component of the {\it s}-process.

\begin{figure}[ht]
\centering\includegraphics[angle=0,keepaspectratio,width=8.5cm]{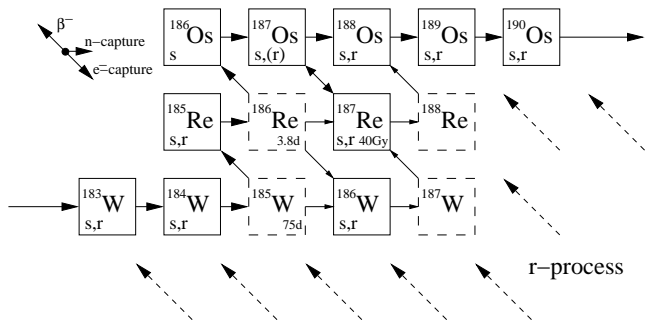}
\caption{\label{fig:fig0} The {\it s}-process path in the W-Re-Os mass region. Unstable nuclei are marked by 
dashed boxes (except $^{187}$Re). The indicated values are laboratory half-lives. However, the half-life of 
$^{187}$Re decreases by 10 orders of magnitude at typical {\it s}-process temperatures of 
$T=3 \times 10^8$\,K and $^{187}$Os becomes unstable\,\cite{Bosc96, Yoko83, Taka83}.}
\end{figure}

Due to its very long half-life ($t_{1/2}=5 \cdot 10^{10}\,{\rm a}$) the 
nucleus $^{187}$Re can be used as a {\it r}-process chronometer\,\cite{Clay64, Brow76}. 
The ratio $N(^{187}{\rm Re}) / N_{\rm c}(^{187}{\rm Os})$ is related to the 
starting point of the {\it r}-process in our galaxy and, hence, to its age. $N$ denotes 
the total and $N_{\rm c}$ the cosmoradiogenic part of the abundance stemming from the 
decay of $^{187}$Re. To extract the cosmoradiogenic 
part of the $^{187}$Os abundance one has to subtract the {\it s}-process 
abundance $N_{\rm s}$ from the total abundance $N$. In Fig.\,\ref{fig:fig0} the {\it s}-process
flow through the W-Re-Os isotopes is shown. The {\it s}-process abundance of $^{187}$Os
can be derived from the abundance of the neighboring {\it s}-only nucleus $^{186}$Os via the
local approximation\,\cite{Clay61}:
\begin{equation}
N_{\rm s}(^{187}{\rm Os})/N_{\rm s}(^{186}{\rm Os}) \approx 
F \,\bar{\sigma}_{\rm n}(^{186}{\rm Os})/\bar{\sigma}_{\rm n}(^{187}{\rm Os}), \label{eq01a}
\end{equation}
where $\bar{\sigma}_{\rm n}$ are the Maxwellian-averaged radiative neutron capture 
cross sections (MACS) from the ground state, and $F$ accounts for the correction 
of the cross section due to neutron capture on thermally excited states in 
$^{187}\rm{Os}$, in particular on the first excited state at $9.75$\,keV. This correction 
factor was first calculated in\,\cite{Woos79} (see\,\cite{Arno84} for discussion). The neutron 
capture cross sections of $^{186}{\rm Os}$ and $^{187}{\rm Os}$ were measured by 
Browne \& Berman\,\cite{Brow76}, Browne, 
Lamaze \& Schroder\,\cite{Brow76b}, by Browne \& Berman\,\cite{Brow81} and 
Winters \& Macklin\,\cite{Wint82}, resulting in an uncertainty of about 20\% for the ratio 
$R=\bar{\sigma}_{\rm n}(^{186}{\rm Os})/\bar{\sigma}_{\rm n}(^{187}{\rm Os})$. 
Recently, these cross sections were measured by the n\_TOF collaboration\,\cite{Mosc04, Kaep05}.

The use of the Re/Os clock is not free of problems. First of all, the half-life 
of $^{187}{\rm Re}$ strongly depends on temperature and $^{187}{\rm Os}$ becomes unstable under
stellar conditions\,\cite{Bosc96, Yoko83, Taka83}. 
Thus, it is necessary to use chemical evolution models of the galaxy\,\cite{Yoko83} 
to include irradiation effects on the abundance ratio $R$. The two branchings at 
$^{185}$W and $^{186}$Re also affect the {\it s}-process abundances in this region. Finally, 
the $N_s \bar{\sigma}_{\rm n}$ correlation Eq.\,(\ref{eq01a}) for the two 
{\it s}-only isotopes $^{186,187}{\rm Os}$ is not fulfilled. This can be caused by two facts. 
Either the branchings are not correctly modeled or the capture cross sections are 
strongly affected by stellar conditions. The branching at $^{185}$W has already
been studied and an overproduction of $^{186}{\rm Os}$ was reported due to the new experimental 
value\,\cite{Sonn03a}. Thus the radiative neutron capture cross section of $^{186}$Re is the 
only relevant cross section in this mass region which is not known experimentally yet. 

In this paper we study the radiative neutron capture cross section of the branching point 
nucleus $^{186}$Re using an indirect method. The unstable nucleus $^{186}$Re decays via $\beta^-$-decay 
to $^{186}$Os or via electron capture to $^{186}$W with a half-life of 
$t_{1/2}=3.7\,{\rm d}$. Due to the fact that neutron capture experiments with 
such short-lived targets are nearly impossible, we choose the inverse
reaction $^{187}$Re($\gamma$,n)$^{186}$Re for our investigation. After neutron emission the $^{186}$Re
nucleus is in the ground state or some excited state and the measured cross section is a sum over several
channels. With maximum excitation energies just above the neutron threshold, only the lowest states can be
reached, e.g., the first excited state in $^{186}$Re at $59\,\rm{\,keV}$. On the other hand, these 
low lying states in the $^{186}$Re nucleus are also thermally populated - however not in the same
proportions - under {\it s}-process conditions and
contribute to the neutron capture cross section. Thus, the $^{187}$Re($\gamma$,n)$^{186}$Re cross section
and the cross section of the inverse reaction $^{186}$Re(n,$\gamma$)$^{187}$Re are related via the 
principle of detailed balance. 

In section \ref{para:par2} we describe our experimental setup. Section \ref{para:par3}
explains the analysis of our data and the results for the $^{187}$Re($\gamma$,n) 
cross section are presented. The results are compared to calculations using the 
NON-SMOKER\,\cite{Raus00a, Raus04} and MOST\,\cite{Gori02, Goriely} codes. Both computer codes are
based on the statistical Hauser-Feshbach model but use different input parameters. In section \ref{para:par3b} 
the implications on the Re/Os clock are briefly discussed.

\section{\label{para:par2} Experimental Setup}

\begin{figure}[ht]
\centering\includegraphics[angle=0,keepaspectratio,width=8.5cm]{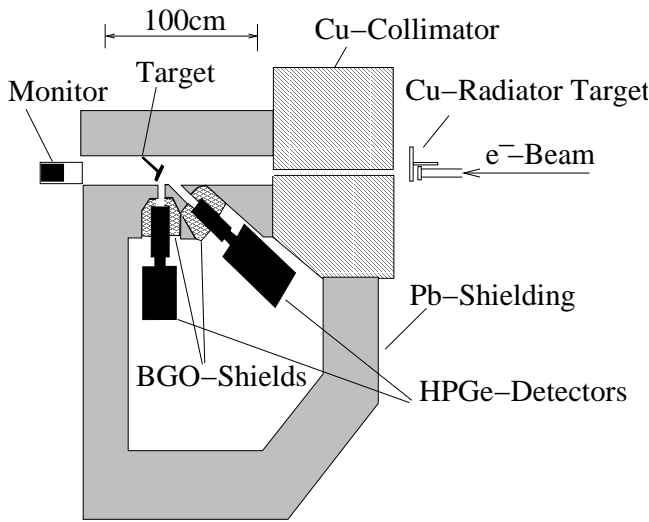}
\caption{\label{fig:fig0c} The real photon setup at the superconducting electron accelerator 
S-DALINAC at Darmstadt University of Technology. The electron beam with a maximum energy of
10\,MeV and maximum intensity of $40\,\mu\rm{A}$ hits a thick copper target and produces
bremsstrahlung. The beam position is monitored with an ionization chamber. The flux
is monitored online with two actively shielded 100\% HPGe detectors using 
the $^{11}$B($\gamma$,$\gamma$') reaction.}
\end{figure}

\begin{figure}[ht]
\centering\includegraphics[angle=0,keepaspectratio,width=8.5cm]{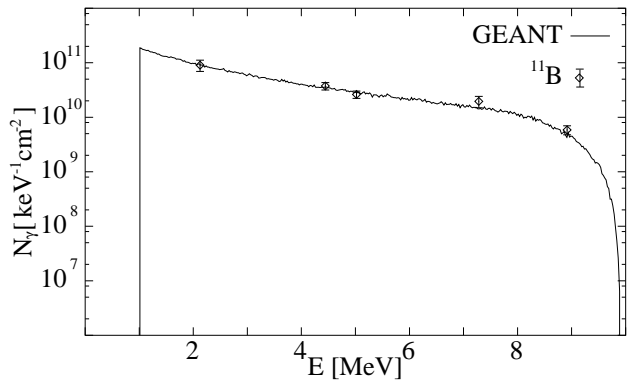}
\caption{\label{fig:fig0b} Adjustment of the simulated spectral photon distribution to
the intensities $N_\gamma$ obtained from the reaction $^{11}$B($\gamma,\gamma$') 
at $E_{\rm max}=9.9\,{\rm MeV}$. The flux is integrated over 23 hours, a typical 
activation time.}
\end{figure}

The $^{187}$Re($\gamma$,n)$^{186}$Re experiment was performed using the photoactivation
technique at the real photon setup\,\cite{Mohr99a} at the superconducting electron accelerator 
S-DALINAC\,\cite{Rich96} at Darmstadt University of Technology (see Fig.\,\ref{fig:fig0c}).  
The monoenergetic electron beam hits a thick
copper radiator ($d \approx 1.4\,{\rm cm}$), where it is completely stopped and converted 
into a continuous spectrum of bremsstrahlung photons with an endpoint energy $E_{\rm max}$. 
The photons are collimated 
and irradiate the target of interest at a distance of $d \approx 150\,{\rm cm}$ 
behind the radiator. 
This leads to a well-defined photon beam with a spectral composition that 
is analyzed in detail, see e.g.\,\cite{Vogt01b}. The $\gamma$-intensity as well as the 
electron current are monitored online in order to control the beam position on the copper
radiator. The targets 
consist of thin metallic rhenium discs ($m \approx 340\,{\rm mg}$, $\varnothing=2\,{\rm cm}$) 
of natural isotopic composition and of two layers of boron ($m \approx 650\,{\rm mg}$ each, 
$\varnothing=2\,{\rm cm}$) embedding the rhenium disc with a sandwich-like structure. 
The incoming photon intensity is normalized using the nuclear resonance fluorescence 
reaction $^{11}$B($\gamma$,$\gamma$')\,\cite{Mohr02b} (see Fig.\,\ref{fig:fig0b}). 
The scattered photons are registered online 
with two HPGe detectors positioned at $90^{\circ}$ and $130^{\circ}$ with respect 
to the beam direction. The detector efficiency $\epsilon$ was measured up to $3.6$\,MeV 
using standard calibration sources. We have used the Monte Carlo code GEANT\,\cite{GEANT} to extrapolate 
the efficiencies up to an energy of $10$\,MeV.

The spectral photon distribution was simulated with the same code and is adjusted 
to the photon intensities obtained from the $^{11}$B($\gamma$,$\gamma$') reaction. 
The result for the case $E_{\rm max}=9.9\,{\rm MeV}$ is shown in Fig.\,\ref{fig:fig0b}. 
A list of all measured energies is shown in Tab.\,\ref{tab:table0b}.
The statistic uncertainty for the photon flux calibration is very small because of the well 
known photo-response of $^{11}$B\,\cite{Ajze90}. The systematic uncertainty includes a $5\%$ uncertainty 
from the detector efficiency extrapolation to higher energies and the uncertainty due to the 
shape of the photon spectrum close to the endpoint energy. This uncertainty is smaller for higher 
endpoint energies because the $\gamma$-transitions at $8.92$\,MeV in the reaction $^{11}$B($\gamma$,$\gamma$') 
can be used for calibration if $E_{\rm max} >9$\,MeV.
The total uncertainty for the photon flux calibration can be estimated to be about 12\,\% to 21\,\% depending 
on the endpoint energy $E_{\rm max}$ (see Tab.\,\ref{tab:table0}).

\begin{figure}[t]
\centering\includegraphics[angle=0,keepaspectratio,width=8.5cm]{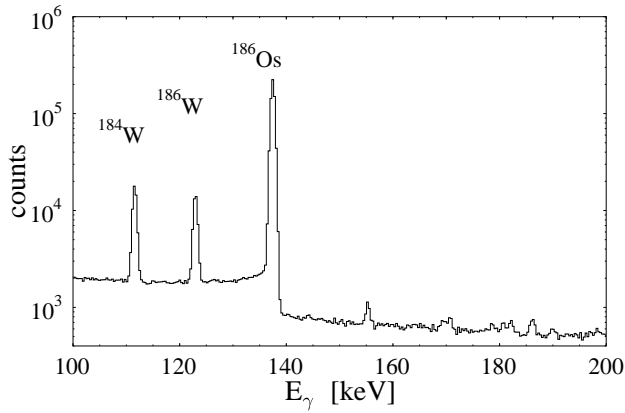}
\caption{\label{fig:fig1} Typical $\gamma$-spectrum measured after photoactivation of $^{\rm nat}$Re. 
The line at 122\,keV and at 137\,keV stem from the $\beta$-decay of $^{186}$Re into $^{186}$W 
and $^{186}$Os, respectively. The line at 111\,keV stems from the decay of $^{184}$Re into $^{184}$W. 
The target was activated for 23 hours at 
an energy of $E_{\rm max}$=9.9\,MeV and the $\gamma$-spectrum was accumulated over a period of 23\,hours 
starting one hour after activation.}
\end{figure}

After activation, the $\gamma$-rays from the decay of the unstable nucleus produced are 
measured offline using a well shielded HPGe detector with an energy 
resolution better than 0.15\,\% and an efficiency of 30\,\% relative to a $3'' \times 3''$ NaI detector. 
The detector efficiency is determined between $60$\,keV and $1.3$\,MeV with standard calibration sources. 
The target geometry and self absorption effects are simulated with a Monte Carlo code \cite{GEANT}. 
A typical spectrum after activation is shown in Fig.\,\ref{fig:fig1}.

\begin{table*}
\caption{\label{tab:table0b} The peak area $Y$ of the two $\gamma$-transitions following the $\beta$-decay 
of $^{186}$Re and the corresponding number of ($\gamma$,n)-reactions $\cal{R}$. The results for the 
transition at $122$\,keV at $E_{\rm max}=7.65$\,MeV are omitted due to low statistics.}
\begin{ruledtabular}
\begin{tabular}{lllll}
$E_{\rm max}$\,[MeV] & Y($122$\,keV)\,[$10^3$] & Y($137$\,keV)\,[$10^3$]  & $\cal{R}$($122$\,keV)\,[$10^6$] 
& $\cal{R}$($137$\,keV)\,[$10^6$]  \\
\hline
9.9 & $35.12 \pm 0.22$ & $605.7 \pm 0.8$ & $287 \pm 14$ &  $286 \pm 14$ \\
9.45 & $17.48 \pm 0.15$ & $301.3 \pm 0.6$ & $148 \pm 8$ &  $147 \pm 7$ \\
9.0 & $9.91 \pm 0.20$ & $167.6 \pm 0.4$ &  $85.0 \pm 4.6$ &  $83.0 \pm 4.2$ \\
8.55 & $3.36 \pm 0.07$ & $58.81 \pm 0.25$ & $33.7 \pm 1.8$ &  $34.1 \pm 1.7$ \\
8.325 & $1.79 \pm 0.05$ & $29.66 \pm 0.18$ & $16.3 \pm 1.0$ &  $15.6 \pm 0.8$ \\
8.1 & $0.87 \pm 0.04$ & $14.98 \pm 0.18$ & $6.86 \pm 0.48$ &  $6.85 \pm 0.35$ \\
7.875 & $0.50 \pm 0.04$ & $7.62 \pm 0.09$ &  $3.24 \pm 0.31$ &  $2.85 \pm 0.15$ \\
7.65 &                 &  $6.82 \pm 0.08$  &                  &  $0.39 \pm 0.02$
\end{tabular}
\end{ruledtabular}
\end{table*}

The peak area $Y$ is directly proportional to the number of ($\gamma$,n)-reactions $\cal{R}$ 
in the target during the activation. Knowing the detector efficiency $\epsilon$ and the 
absolute $\gamma$-branching of the transition, the factor of proportionality can be directly calculated 
for each run using the law of exponential decay. The number of ($\gamma$,n)-reactions $\cal{R}$ 
is proportional to the energy-integrated cross section $I_{\rm \sigma}$,
\begin{equation}
{\cal R} = n_t\,I_{\rm \sigma} = n_t\,\int_{S_{\rm n}}^{E_{\rm max}} 
N_{\gamma}(E,E_{\rm max}) \sigma (E) \,dE \label{eq1}
\end{equation}
where $n_t$ is the number of $^{187}$Re atoms, $S_{\rm n}$ is the neutron threshold, 
$E_{\rm max}$ is the energy of the electron beam and, hence, the endpoint energy of the 
spectral density distribution 
$N_{\gamma}(E,E_{\rm max})$ of the bremsstrahlung photons, and $\sigma (E)$ is the
$^{187}$Re($\gamma$,n)$^{186}$Re cross section. We used the strong transitions at 
$137$ and $122$\,keV, respectively, for our analysis.

The uncertainty of $I_\sigma$ is about 5\,\% to 6\,\%. The statistical uncertainty 
is small due to the high sensitivity of the photoactivation technique (see Fig.\,\ref{fig:fig1}). 
The uncertainty of $I_\sigma$ is dominated by the 5\,\% uncertainty stemming from the simulation of the
detector efficiency including self-absorption effects in the target. The peak areas 
$Y$ and the number of ($\gamma$,n)-reactions $\cal{R}$ are summarized in Tab.\,\ref{tab:table0b}. 
The different components of the uncertainties as well as the total uncertainty are summarized in 
Tab.\,\ref{tab:table0} for each endpoint energy $E_{\rm max}$.

\begin{table*}
\caption{\label{tab:table0} Summary of experimental uncertainties for the different endpoint energies $E_{\rm max}$.
The details are discussed in the text.}
\begin{ruledtabular}
\begin{tabular}{lllllll}
& \multicolumn{3}{l}{$\gamma$-intensity} & \multicolumn{2}{l}{number of ($\gamma$,n)-reactions} & \\
$E_{\rm max}$\,[keV] & stat.\,[\%] & sys.\,[\%] & shape\,[\%] & stat.\,[\%] & sys.\,[\%] & Total\,[\%]\\
\hline
9900 & 1.0 & 9.8 & 8 & 0.1 & 5 & 13.6\\
9450 & 1.4 & 8.6 & 8 & 0.2 & 5 & 12.8\\
9000 & 1.2 & 10.7 & 11 & 0.2 & 5 & 16.2\\
8550 & 1.3 & 12.9 & 11 & 0.4 & 5 & 17.7\\
8325 & 1.5 & 10.7 & 11 & 0.6 & 5 & 16.2\\
8100 & 2.0 & 14.1 & 14 & 0.8 & 5 & 20.6\\
7875 & 1.8 & 14.9 & 14 & 1.2 & 5 & 21.2\\
7650 & 2.0 & 13.0 & 14 & 1.2 & 5 & 19.9
\end{tabular}
\end{ruledtabular}
\end{table*}

\section{\label{para:par3} Analysis of the data}

The peak area in the activation spectrum is only proportional to $I_\sigma$, thus, we cannot determine the
cross section directly. Assuming a certain shape of the cross section, one can verify and normalize this
assumption by measuring at several endpoint energies because the centroid of the integrand in 
Eq.\,(\ref{eq1}) changes for each endpoint energy.

\begin{figure}[ht]
\centering\includegraphics[angle=0,keepaspectratio,width=8.5cm]{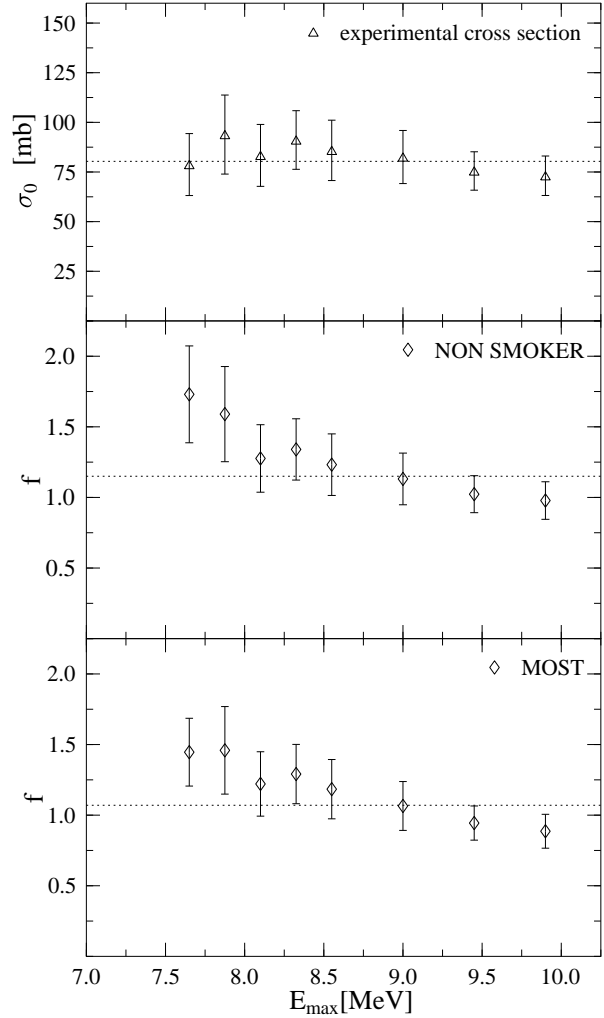}
\caption{\label{fig:fig3} The normalization factors $\sigma_0$ of Eq.\,(\ref{eq2}), 
alternatively $f$ of Eq.\,(\ref{eq6})
are shown as a function of the endpoint energy $E_{\rm max}$ of the photon spectrum for our experimental cross 
section and for the two theoretical predicted cross sections. Both model predictions seem to 
underestimate the cross section slightly in the vicinity of the reaction threshold.}
\end{figure}

\begin{figure}[ht]
\centering\includegraphics[angle=0,keepaspectratio,width=8.65cm]{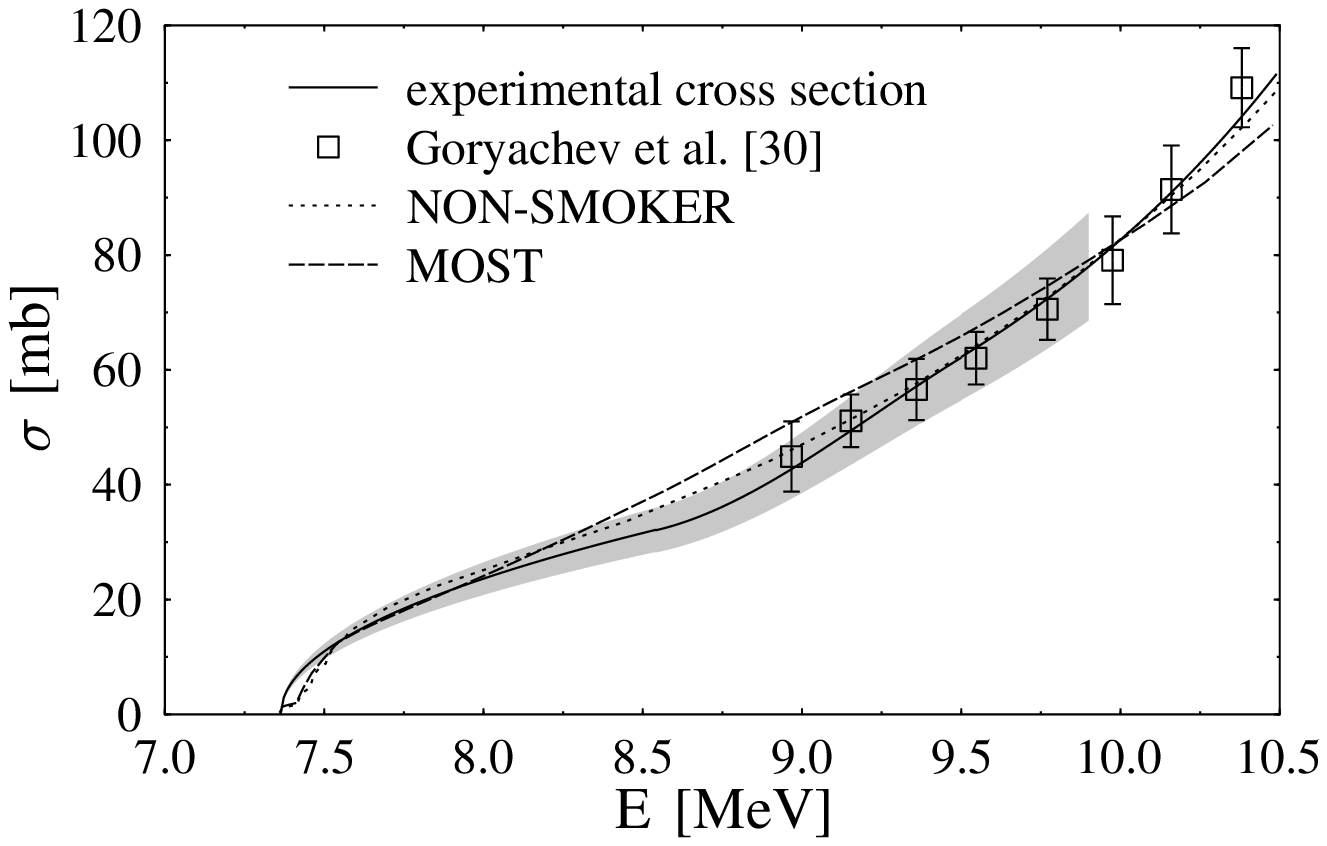}
\caption{\label{fig:fig2} Comparison of the experimental $^{187}$Re($\gamma$,n) cross
section to the calculated cross sections from the  MOST and NON-SMOKER code. 
Additionally, the data points extracted from Goryachev {\it et al.}\,\cite{Gory73} 
are plotted. The shaded area represents the uncertainty of our experimental results.}
\end{figure}

In case of pure s-wave neutron emission, the ($\gamma$,n) cross section can be 
parametrized as\,\cite{Beth37}
\begin{equation}
\sigma (E) = \sigma_0 \sqrt{\frac{E-S_{\rm n}}{S_{\rm n}}} \label{eq2}
\end{equation}
for the energy region close above the neutron threshold energy $S_{\rm n}(^{187}{\rm Re})=7.363\,{\rm MeV}$.
The combination of Eq.\,(\ref{eq2}) with Eq.\,(\ref{eq1})  yields 
the normalization factor $\sigma_0$.  The normalization factors shown in Fig.\,\ref{fig:fig3} 
as a function of endpoint energy should be constant. We have measured the integrated 
cross section using bremsstrahlung with endpoint energies of 7.65, 7.875, 
8.1, 8.325, and 8.55\,MeV, respectively. For $S_{\rm n} < E < 8.55 \,{\rm MeV}$ we obtain:
\begin{equation}
\sigma (E)=(80.4 \pm 9.6) \,{\rm mb} \,\sqrt{\frac{E-S_{\rm n}}{S_{\rm n}}}\label{eq2a}. 
\end{equation}
Between 9.5 and 19\,MeV, the ($\gamma$,n) cross section can be parametrized 
by a superposition of two Lorentzians\,\cite{Gory73}. Between 8.55 and 9.5\,MeV,
we interpolate the cross section with a third order polynomial. The requirement 
that the cross section as well as its first derivative are continuous, 
determines the four parameters of the  polynomial:
\begin{eqnarray}
\left(\frac{\sigma(E)}{\rm mb}\right) & = &-17 \left(\frac{E}{\rm MeV}\right)^3
+4.7\times10^2 \left(\frac{E}{\rm MeV}\right)^2 \nonumber \\ 
&-& 4.3\times10^3\left(\frac{E}{\rm MeV}\right)+1.3\times 10^4 \label{eq2b}
\end{eqnarray}
The parametrization of the cross section for the energy range between $S_{\rm n}$ and 10.5\,MeV 
is shown in Fig.\,\ref{fig:fig2}. The measurements performed at $E_{\rm max}= 9.0$, 9.45 and 
9.9\,MeV  verify this parametrization up to the low energy tail 
of the giant dipole resonance. Our data fit nicely with an older experiment\,\cite{Gory73}, 
in which the $^{187}$Re($\gamma$,n) cross section was measured at higher 
energies around the giant dipole resonance.

Alternatively  one can start with a cross section $\sigma_{\rm th}$ calculated with the NON-SMOKER 
or MOST code. A normalization factor $f$ is introduced for absolute calibration:
\begin{equation}\label{eq6}
I_{\rm \sigma} = f \cdot \int_{S_{\rm n}}^{E_{\rm max}} \sigma_{\rm th} (E) N_{\gamma}(E,E_{\rm max}) \,dE
\end{equation}

The normalization factor $f$ needs to be energy independent  
presuming that the predicted shape of the cross section $\sigma_{\rm th}$ 
is correct. Additionally, if the absolute value of the calculated
cross section is correct, $f$ should be close to $1.0$.
The normalization factors $f$ are determined to be
\begin{eqnarray}
\label{eq7} f_{\rm NONS}&=&1.15 \pm 0.31 \\
\label{eq8} f_{\rm MOST}&=&1.07 \pm 0.28
\end{eqnarray}
The normalization factors as a function of the endpoint energy $E_{\rm max}$ are shown in 
Fig.\,\ref{fig:fig3} and the normalized cross sections are shown in Fig.\,\ref{fig:fig2}.
The agreement 
between the two calculations and our experimental cross section is remarkably good. The small deviations
in the threshold region result either from experimental uncertainties regarding the precise shape
of the bremsstrahlung close to the endpoint energy $E_{\rm max}$ or from numerical problems occurring in
the theoretical calculation.

\section{\label{para:par3b} Implications for the Re/Os clock}

The current uncertainty of the time duration of nucleosynthesis obtained from the Re/Os clock
is about 2.3\,Gyr\,\cite{Meng01}. The main source of uncertainty with regard to nuclear physics aspects of 
the clock are presently the neutron capture cross sections. 

The Maxwellian-averaged neutron capture cross sections in Tab.\,\ref{tab:table1} 
are derived using the MOST and NON-SMOKER code. The ($\gamma$,n) cross sections are in good agreement 
with the present experimental results (see Fig.\,\ref{fig:fig2}). 
However, the neutron capture cross sections of the inverse reaction differ 
by a factor of 2.4.

\begin{table}
\caption{\label{tab:table1} The MACS calculated with 
an updated version of NON-SMOKER\,\cite{Raus04} and MOST\,\cite{Goriely} for a typical {\it s}-process 
temperature of $kT=30$\,keV. The stellar enhancement factor as well as the adopted values of the MACS 
were taken from Bao {\it et al.}\,\cite{Bao00}.}
\begin{ruledtabular}
\begin{tabular}{lcr}
 & $\sigma_{\rm GS}$ [mb] & $\sigma_{\rm thermal}$ [mb] \\
\hline
NON-SMOKER\,\cite{Raus04} & 1485 & 1546 \\
MOST\,\cite{Goriely} & 616 & 623 \\
Ref.\,\cite{Bao00} & $1550 \pm 250$ & $1615 \pm 260 $\\
\end{tabular}
\end{ruledtabular}
\end{table}

Taking the value recommended by Bao {\it et al.}\,\cite{Bao00} the branching ratio between 
$\beta$-decay and neutron capture is $R=\lambda_n / \lambda_\beta=5.4 \times 10^{-4}$. Considering the values 
from the NON-SMOKER and MOST calculations, one obtains $R_{\rm NONS}=5.2 \times 10^{-4}$ and 
$R_{\rm MOST}=2.1 \times 10^{-4}$, respectively. Both models predict a (n,$\gamma$)-cross 
section which is smaller than the value recommended 
in\,\cite{Bao00}. This would lead to a further enhanced production
of $^{186}$Os. Nevertheless due to the small branching the abundance distribution 
of the elements in this mass region is not changed significantly. 

\section{\label{para:par4} Summary and conclusion}

We have measured the $^{187}$Re($\gamma$,n) cross section just above 
the reaction threshold. The cross sections calculated within the statistical 
model using the computer codes NON-SMOKER and MOST are in good agreement 
with our experimental data. Even though the predictions for the photodisintegration
cross sections are nearly identical, the MACS for the neutron capture reactions differ by a 
factor of 2.4. At this point, further theoretical investigations are necessary. 

The MACS calculated with the updated NON-SMOKER code is close to the
value given in\,\cite{Bao00}. Thus, the overproduction of $^{186}$Os reported 
in\,\cite{Sonn03a} remains. The MACS calculated with the MOST code is significantly 
lower and, therefore, a further increase of the $^{186}$Os production is predicted. For a 
more quantitative statement, complex network calculations are mandatory. 
 
The disagreement between the two models may have its origin in the lack of 
precise nuclear data that enter into both models. This shows again the need for high 
precision mass measurements, E1-strength distribution studies, and the study of 
nucleon-nucleus optical potentials.

The fact that both model predictions are in good agreement with our data but could not 
reduce the overproduction of $^{186}$Os supports the idea that the adopted 
value of the $^{186}$Os(n,$\gamma$) cross section\,\cite{Bao00} is too small. 
Very recent results from an experiment at the n-TOF facility at CERN\,\cite{Mosc04} and at the 
Forschungszentrum Karlsruhe\,\cite{Kaep05} point in this direction as well.\\

\begin{acknowledgments}
We thank S.\ Goriely and T.\ Rauscher for performing the  
calculations with the MOST and NON-SMOKER code, respectively. 
We thank the S-DALINAC group around H.-D.\ Gr\"af for their support
during the experiment and the members of our group, especially M.\ Babilon, 
W.\ Bayer, K.\ Lindenberg, D.\ Savran, and S.\ Volz for their help during 
the beam time. We thank F.\ K\"appeler, T.\ Shizuma  and H.\ Utsunomiya for
encouraging discussions. We thank F.\ K\"appeler for a careful reading of the manuscript.
We thank P.\ Mohr for initiating this experiment and 
for discussions. This work was supported by the Deutsche Forschungsgemeinschaft 
under contract SFB 634.
\end{acknowledgments}

\bibliographystyle{/userx/users/mueller/pub/bib/bibtex/apsrev}
\bibliography{/userx/users/mueller/pub/bib/bibtex/english}

\end{document}